\documentclass[pre,twocolumn,aps,floatfix,10pt]{revtex4}
\usepackage{amsmath}
\usepackage{amsfonts}
\usepackage{amssymb}
\usepackage{graphicx}
\usepackage{fontenc}
\usepackage{psfrag,layout}
\usepackage{upgreek}

\begin{document}
\title{Lennard-Jones systems near solid walls: Computing interfacial 
free energies from molecular simulation methods}
\author{Ronald Benjamin and J{\"u}rgen Horbach}
\affiliation{Institut f{\"u}r Theoretische Physik II, 
Heinrich-Heine-Universit{\"a}t D{\"u}sseldorf,
Universit\"atsstra\ss e 1, 40225 D{\"u}sseldorf, Germany}

\begin{abstract}
Different computational techniques in combination with molecular dynamics
computer simulation are used to to determine the wall-liquid and the
wall-crystal interfacial free energies of a modified Lennard-Jones
(LJ) system in contact with a solid wall.  Two different kinds of solid
walls are considered: a flat structureless wall and a structured wall
consisting of an ideal crystal with the particles rigidly attached
to fcc lattice sites.  Interfacial free energies are determined by a
thermodynamic integration scheme, the anisotropy of the pressure tensor,
the non-equilibrium work method based on Bennett acceptance criteria,
and a method using Cahn's adsorption equations based on the interfacial
thermodynamics of Gibbs.  For the flat wall, interfacial free energies as
a function of different densities of the LJ liquid and as a function of
temperature along the coexistence curve are calculated. In case of a structured wall,
the interaction strength between the wall and the LJ system and the
lattice constant of the structured wall are varied.  Using the values
of the wall-liquid and wall-crystal interfacial energies along with
the value for the crystal-liquid interfacial free energy determined
previously for the same system by the ``cleaving potential method'', we obtain
the contact angle as a function of various parameters; in particular
the conditions are found under which partial wetting occurs.
\end{abstract}

\maketitle

\section{Introduction}
\label{sec_intro}

Interfacial free energies between a crystal or liquid in contact with
a solid wall are a key determinant of several  interfacial phenomena
such as wetting~\cite{degennes85}, adhesion and heterogeneous
nucleation~\cite{zettlemoyer69,abraham74,kashchiev00}.
However, they are difficult to determine in experiments
and very few results are available in the literature~(see
e.g.~\cite{adamson97,navascues79,howe97}).  Therefore, molecular
simulation techniques such as Molecular Dynamics (MD) and Monte Carlo (MC)
\cite{frenkel-smit02,landau-binder00,allen-tildesley87} play an important
role to elucidate interfacial phenomena.  Since molecular simulations
can be used to determine the same thermodynamic quantity via various
computational strategies, reliable estimates of the desired quantities
are obtained by comparing the results obtained from the different methods.
In this work, our objective is to employ several different computational
techniques for obtaining reliable estimates of the wall-liquid and
wall-crystal interfacial free energies, in the following denoted by
$\gamma_{\text{wl}}$ and $\gamma_{\text{wc}}$, respectively.

In order to calculate $\gamma_{\text{wl}}$, mechanical techniques
can be employed to compute the interfacial tension from the
pressure anisotropy (PA)~\cite{kirkwood49}.  Several authors have
used this technique to calculate $\gamma_{\rm wl}$ for both hard-sphere
\cite{henderson84,courtemanche93,miguel06} and Lennard-Jones (LJ) systems
\cite{sikkenk88,bakker89,nijmeijer90,bruin91,crevecoeur95,tang95,varnik00}.
Since, in general, the interfacial tension is not equal to the interfacial
free energy~\cite{shuttleworth50}, except in case of a liquid in contact
with a static wall, the mechanical approach fails in situations involving
liquid in contact with a wall consisting of a ``fully interacting
solid phase''~\cite{laird10} or for a crystal in contact with a solid
wall~\cite{tiller91}. Another drawback is the massive computational
effort needed to obtain reliable results, since the interfacial tension
being the difference of the normal and tangential pressure profiles is
very sensitive to small numerical errors~\cite{deb10,deb11}.

In contrast to the PA method, thermodynamic integration
(TI) techniques~\cite{frenkel-smit02,benjamin-horbach2012} are
applicable to a wider class of systems including systems where
the liquid interacts with an elastic wall or for wall-crystal
interfaces.  Various TI schemes have been proposed for the
calculation of $\gamma_{\rm wl}$ and $\gamma_{\rm wc}$ and applied
to systems modelled by hard-sphere, LJ or other soft interaction
potentials~\cite{benjamin-horbach2012,heni99,fort-djik06,laird07,leroy09}.
While the TI method is simple to understand and easy to implement in
MD or MC simulations, care has to be taken that the switching protocol
that transforms the initial to the final state is reversible and does
not lead to any hysteresis. As a result, the computational load can
be considerable.

A less computationally expensive method, which does not require the
constraint of a reversible switching protocol, is the  non-equilibrium
work method. In 1997, Jarzynski showed that one can extract the
equilibrium free energy difference between two states from an exponential
average of the non-equilibrium work performed on the system during a
transformation from one state to another~\cite{jarzynski1997}.  However,
to obtain an accurate estimate of the free energy difference using the
Jarzynski equality (JE), the tail of the work probability distribution
must be well sampled, which is possible only if an extremely large
number of independent runs are performed~\cite{jarzynski06}. Therefore,
in general, a direct application of the Jarzynski equality leads to a
significant bias and variance in the free energy estimates~\cite{fox03}.
Shirts {\it et al.}~\cite{shirts-pande2005} demonstrated that one
can overcome this disadvantage by carrying out both forward and
backward transformations between the two states and then estimating the
equilibrium free energy difference using the Bennett acceptance ratio
(BAR)~\cite{bennett1976}.  The BAR method has been applied to obtain
the crystal-liquid interfacial free energy for both hard sphere and LJ
systems~\cite{davidchack2010,musong2006}. However, to the best of our
knowledge there have been no studies where non-equilibrium work approaches
have been used to compute $\gamma_{\text {wl}}$ or $\gamma_{\text{wc}}$.

If interfacial free energies are needed at many points along
a coexistence curve, say, the TI and BAR methods would both be
computationally expensive.  Recently, Laird and others have obtained
$\gamma_{\text{wl}}$ for a hard-sphere system by integrating along a
pressure curve~\cite{laird10} and the crystal-liquid interfacial free
energy for LJ systems as a function of temperature along the coexistence
curve~\cite{laird09} using  the ``Gibbs-Cahn integration" method.  This
technique is based on adsorption equations derived by Cahn~\cite{cahn79}
to extend Gibbs's interfacial thermodynamics~\cite{gibbs57}.  The
``Gibbs-Cahn integration" (GC) method requires that the values for the
interfacial free energy is already known at one point along the curve,
then at the remaining points the free energies can be determined by
integrating along the curve.  While, Laird and Davidchack computed
$\gamma_{\text{wl}}$ for hard spheres in contact with a hard wall,
there have been no corresponding studies for LJ systems.

Although, the above described  methods can be employed to determine
$\gamma_{\text{wl}}$ and $\gamma_{\text{wc}}$, {\it a priori} it is not
known which technique yields the most accurate results.  To investigate
this issue, we use different computational methods to arrive at reliable
estimates for $\gamma_{\text{wl}}$ and $\gamma_{\text{wc}}$ and also
discuss the computational efficiency of each method.  To this end, we
compute $\gamma_{\text{wc}}$ and $\gamma_{\text{wl}}$ for a LJ system in
contact with solid walls from non-equilibrium work measurements using the
JE and BAR methods and compare the results with a TI scheme devised in our
earlier work~\cite{benjamin-horbach2012}. The interfacial free energies
are computed for both a bulk LJ liquid and a bulk LJ face centred cubic
crystal oriented along the (111) direction. We consider two kinds of solid
walls: one is a flat structureless wall and the other one is a structured
wall consisting of particles rigidly attached to fcc lattice sites with
the (111) orientation of the wall in contact with the system.  In this
work, $\gamma_{\text{wl}}$ and $\gamma_{\text{wc}}$ are obtained by our
TI scheme and non-equilibrium work methods; the wall-liquid interfacial
free energy, $\gamma_{\text{wl}}$ is also determined via the GC method
as well as the PA technique. Results from the four different approaches
are compared with each other.

If in addition to the wall-liquid and wall-crystal interfacial free
energies, one has knowledge about the crystal-liquid interfacial free
energy, $\gamma_{\text{cl}}$, a direct determination of the degree of
wetting of a solid surface by a substance can be obtained from Young's
equation~\cite{young1805},
\begin{equation}
\gamma_{\text{wc}}+\gamma_{\text{cl}}
\cos \theta_{\text{c}}=\gamma_{\text{wl}}\,
\label{eq:young}
\end{equation}
with $\theta_{\rm c}$ the contact angle.  The contact angle controls
whether crystal nucleation from the bulk liquid occurs preferably
via homogeneous nucleation corresponding to complete drying
($\theta_{\rm c} = 180^\circ$) or at the wall via heterogeneous nucleation
corresponding to partial wetting ($0^\circ <\theta_{\rm c} < 90^\circ$) or
partial drying ($90^\circ <\theta < 180^\circ$) scenarios.  However,
to obtain $\theta_{\rm c}$ from Eq.~(\ref{eq:young}), the value for
$\gamma_{\text{cl}}$ is also needed. Therefore, we carried out simulations
using a form of the LJ potential first introduced by Broughton and
Gilmer~\cite{broughton-gilmer86,davidchack-laird03,laird09,asano-fuchizaki09,asano-fuchizaki12},
for which $\gamma_{\text{cl}}$ has already been determined using the
``cleaving potential'' approach~\cite{davidchack-laird03} as well as the
Gibbs Cahn integration technique~\cite{laird09}.

In the following, we introduce the details of the model potentials
considered in this work (Sec.~\ref{sec_mp}), then describe the
non-equilibrium work method and the Gibb's Cahn integration
technique (Sec.~\ref{sec_calc}) and provide the main details of
the simulation (Sec.~\ref{sec_sim}). Then, we present the results
(Sec.~\ref{sect:results}) and finally draw some conclusions
(Sec.~\ref{sec_conc}).

\section{Model Potential}
\label{sec_mp}
We consider a system consisting of $N$ identical particles, each of mass
$m$, interacting with each other via a LJ potential first introduced by
Broughton and Gilmer~\cite{broughton-gilmer86}. For two particles $i$ and
$j$ separated by a distance $r$, this interaction potential is defined by
\begin{equation}
u(r) = \left\{ \begin{array}{lll}
      \phi(r)+C_{1} & \mbox{for} & 0 < r \leq  2.3 \sigma \\ 
      C_{2} (\frac{\sigma}{r})^{12} + C_{3} (\frac{\sigma}{r})^{6} & \\
      + C_{4}(\frac{r}{\sigma})^{2} + C_{5} & \mbox{for} & 2.3\sigma < r < 2.5 \sigma \\
      0 & \mbox{for} & r> r^{\text{c}}
        \end{array}\right.
\label{eq:ur}
\end{equation}
with
\begin{equation}
\phi(r) =
4\epsilon \left[ \left(\frac{\sigma}{r}\right)^{12}
- \left(\frac{\sigma}{r}\right)^{6} \right].
\label{eq:lj}
\end{equation}
In the following, energies and lengths are given in units of the
parameters $\epsilon$ and $\sigma$, respectively. The cut-off distance
$r^{c}$ is set to $r^{c}=2.5\sigma$.  The constants in Eq.~(\ref{eq:ur})
are given by $C_{1}=0.016132\,\epsilon$, $C_{2}=3136.6\,\epsilon$,
$C_{3}=-68.069\,\epsilon$, $C_{4}=-0.083312\,\epsilon$, and
$C_{5}=0.74689\,\epsilon$.

To model the interaction of the LJ system with the structured wall
we choose the purely repulsive Weeks-Chandler-Andersen (WCA) potential,
\begin{equation}
u_{\text{sw}}(r)=
4 \epsilon_{\rm sw}
\left[\left( \frac{\sigma_{\text{sw}}}{r} \right)^{12}- 
      \left( \frac{\sigma_{\text{sw}}}{r} \right)^{6} +\frac{1}{4}\right]
\label{eq:uswallwca}
\end{equation}
for $0 < r \leq 2^{1/6}\sigma_{\rm sw}$ and 0 otherwise where $\sigma_{\rm
sw}$ is set to $\sigma_{sw} = 0.941\sigma$. Here, $r$ represents
the distance between a LJ particle in the bulk and a wall particle.
The structured wall was constructed from an ideal fcc crystal with the
(111) orientation along the $z$ direction and with an integer number
of unit cells that fit exactly into the simulation cell.  The reason
for choosing $u_{\text{sw}}(r)$ different from $u(r)$ was to obtain a
large parameter range with respect to the energy scale $\epsilon_{\rm
sw}$ and the lattice constant of the wall $a_{sw}$ [where $a_{sw}=(4/\rho_{sw})^{(1/3)}$, $\rho_{\text{sw}}$ being
the density of the structured wall], providing the possibility of
incomplete wetting conditions. In the case of $u_{\text{sw}}(r)=u(r)$,
complete wetting would be more likely than partial wetting, even for
moderate interaction strengths between the wall and the LJ system.

The $N$ identical LJ particles are enclosed within a simulation cell of
size $L_{\text{x}}\times L_{\text{y}} \times L_{\text{z}}$, with periodic
boundary conditions in the $x$ and $y$ directions. Our simulations are
carried out in the $NVT$ ensemble with the total density $\rho=N/V$
kept constant. Along the $z$ direction the particles are confined by
the walls such that there are two planar wall-liquid (or wall-crystal)
interfaces with a total area of $A=2L_{\text{x}}L_{\text{y}}$ and with
the Gibbs dividing surface located at the surface of the wall.  The width
of the structured wall is chosen large enough to avoid LJ particles on
opposite sides of the wall from interacting with each other since the
determination of interfacial free energy is built on the assumption of
two independent wall-liquid (or wall-crystal) interfaces.

In our earlier work~\cite{benjamin-horbach2012} on computing $\gamma_{\rm
wl}$ and $\gamma_{\rm wc}$ for structured walls via TI, we adopted
a scheme which consists of two steps. First, a bulk LJ system with
periodic boundary conditions is transformed into a state where the LJ
system is in contact with flat walls on either side in the $z$ direction.
Then, in the second step, the flat walls are reversibly transformed into
structured walls.  The structureless flat wall (fw) was also taken to
be a purely repulsive potential interacting along the $z$ direction with
the LJ particles and is described by a WCA potential,
\begin{equation}
u_{\text{fw}}(z)=
 4\epsilon \left[ \left(\frac{\sigma}{z} \right)^{12}
                 -\left(\frac{\sigma}{z} \right)^{6}
 +\frac{1}{4}\right] \times w(z)
\label{eq:ufwallwca}
\end{equation}
for $0 < z \leq z_{\text{cw}}=2^{1/6}\sigma$ and zero otherwise. Note
that for $u_{\text{fw}}(z)$ the parameters $\epsilon$ and $\sigma$ are
identical to those chosen for $u(r)$.  The variable $z=z_{i}-Z$ denotes
the distance between a particle $i$ at $z_{i}$ to one of the flat walls
at $Z=z_{\rm b}$ or $Z=z_{\rm t}$.  The function $w(z)$ ensures that
$u_{\text{fw}}(z)$ goes smoothly to zero at $z=z_{\text{cw}}$ and is
given by
\begin{equation}
 w(z)=\frac{(z-z_{cw})^{4}}{h^{4}+(z-z_{\text{cw}})^4},
\end{equation}
with the dimensionless parameter $h=0.005$.  Below, results for
$\gamma_{\rm wl}$ and $\gamma_{\rm wc}$ for both the flat and structured
wall cases are presented.

\section{Methods}
\label{sec_calc}
When a system is transformed from one equilibrium state to another,
the total work $\langle W \rangle$ performed on the system is larger
than the free energy difference, $\Delta F$, between the states,
\begin{equation}
\label{eq:seclaw}
\langle W \rangle  \geq \Delta F .
\end{equation}
This inequality holds because some of the work is always dissipated as
heat, i.e.
\begin{equation}
\langle W^{\rm diss} \rangle = \langle W \rangle -\Delta F \geq 0
\end{equation}
with $W^{\rm diss}$ the dissipated work.  The equality $\langle W \rangle
= \Delta F$ holds when the process is reversible such that the dissipated
work goes to zero.  This reversible work is essentially what one computes
in the TI method where one changes the switching parameter infinitely
slowly such that the transformation is reversible.  As a result, the TI
method yields the free energy difference between a desired state and some
reference state~\cite{frenkel-smit02}. Usually a parameter which couples
to the interaction potential is gradually changed such that the reference
state is reversibly transformed into the final state of interest.

To calculate the interfacial free energy of the LJ system in contact
with a structured wall, the TI scheme is carried out in two steps. In the 
first step, a bulk LJ system without walls and periodic boundary conditions 
in all directions is reversibly transformed into a LJ system in contact with 
a structureless flat wall along the $z$ direction.  In the second step, the 
flat wall interacting with the LJ system is reversibly transformed into a 
structured wall. Calculating the total free energy difference $\Delta F$ combined 
from the two steps yields  the required interfacial free energy, defined as
\begin{equation}
\label{eq:defife}
 \gamma = \frac{F_{\text{system}} - F_{\text{bulk}}}{A}=\frac{\Delta F}{A}
\end{equation}
with $F_{\text{system}}$ and $F_{\text{bulk}}$ representing the Helmholtz
free energies of the inhomogeneous system and the bulk phase of the
system, respectively. For details we refer the reader to our earlier
work~\cite{benjamin-horbach2012}. In both steps of our TI scheme the
transformation from one equilibrium state to another is carried out
by gradually changing a switching parameter $\lambda$, which couples
directly to the interaction potential.

Although the TI method yields reliable values for interfacial free
energies, it is computationally expensive so as to ensure reversibility.
In principle, the switching protocol needs to be infinitely slow when
computing the free energy difference from a sequential change of the
switching parameter or by carrying out simultaneous simulations at many
independent values of the switching parameter between the initial and
final states, so as to prevent any hysteresis or numerical integration
errors resulting from a non-smooth thermodynamic integrand. Besides,
in many cases a reversible TI path might be difficult to construct.
A less computationally expensive method, which is independent of the
reversibility of the switching protocol is the non-equilibrium work
method.

\subsection{$\gamma$ from non-equilibrium work methods.}
Equation~(\ref{eq:seclaw}) shows that the free energy difference
between two equilibrium states can only be obtained from
a reversible transformation between them.  However, in 1997
Jarzynski~\cite{jarzynski1997} proposed an equality, now known as the
Jarzynski equality, that allows to extract free energy differences
between two states even if the transformation from the initial to the
final state proceeds via a non-equilibrium process.

The Jarzynski equality (JE) can be written as
\begin{equation}
 \langle \exp(-\beta W) \rangle = \exp (-\beta \Delta F)
\end{equation}
with $\beta = 1/(k_B T)$ ($k_B$: Boltzmann constant, $T$: temperature) 
and the work
\begin{equation}
W = \int_{t=0}^{t_{s}}
    \frac{\partial H(\lambda)}{\partial \lambda} \dot{\lambda} dt .
\label{eq:w}
\end{equation}
Here, the Hamiltonian $H(\lambda)$ [see Eq.~(7) of
Ref.~\cite{benjamin-horbach2012} and accompanying text for details
regarding the Hamiltonian] depends on the parameter $\lambda$ that
provides the switching from state to state; $\dot{\lambda}$ is the rate
at which the parameter $\lambda$ is changed to transform one state to
another. The integration in Eq.~(\ref{eq:w}) starts from an equilibrium
state at time $t=0$ to a final state at the switching time $t=t_{\rm s}$.
In a simulation, the free energy difference between the initial and
the finite state is estimated by
\begin{equation}
\Delta F_{\rm JE} = -k_{B}T \ln \sum_{i=1}^{N_{\rm runs}} \exp(-\beta W_i)
\label{eq:jarest}
\end{equation}
with $N_{\rm runs}$ the number of independent runs and $W_i$ the value
of the work, obtained from the $i$'th run. The estimate $\Delta F_{\rm
JE}$ approaches the exact free energy difference $\Delta F$ for $N_{\rm
runs} \to \infty$.

For slow transformations ($t_{\rm s} \rightarrow \infty$), the system is in
quasi static equilibrium at each instant throughout the transformation
and the switching process reduces to a conventional TI. However, for
finite switching times $t_{\rm s}$, the system is in a non-equilibrium state
and while the JE holds irrespective of the duration $t_{\rm s}$ over which
the switching parameter is varied, its practical utility is limited
since the exponential work average depends on how well the tail of the
work probability distribution with work values that are less than the
equilibrium free energy difference are sampled~\cite{jarzynski06}. As a
result, the JE is a practical tool to obtain free energy differences
only for slow switching processes where the system remains close
to equilibrium at each instant during the transformation.  In such a
situation, the average work converges to $\Delta F$ within a reasonable
number of independent runs.  However, for fast switching times, the
exponentially averaged work exhibits a huge bias in comparison to the
equilibrium free energy difference and is also accompanied by a large
standard deviation~\cite{fox03}.

Shirts {\it et al.}~\cite{shirts-pande2005} showed that a good convergence
toward $\Delta F$ can be obtained by carrying out simulations with both
forward as well as reverse switching runs. In the reverse runs, the
final state is transformed into the initial state by following a mirror
image of the forward switching protocol. Using the Bennett acceptance
ratio (BAR) method~\cite{bennett1976}, the free energy difference is
finally obtained from the following transcendental equation,
\begin{equation}
\sum_{i=1}^{N_{F}} \frac{1} {1+\exp [ \frac{W - \Delta F}{k_BT} ] }  -
\sum_{i=1}^{N_{R}} \frac{1} {1+\exp [-\frac{W - \Delta F}{k_BT} ] } = 0
\label{eq:BAR}
\end{equation}
that can be solved via the Newton-Raphson method~\cite{numrecipe}.
In Eq.~(\ref{eq:BAR}), $F$ and $R$ represent respectively the
forward and reverse switching processes and $N_{\rm F}$ and $N_{\rm R}$
refer respectively to the number of forward and reverse switching
trajectories. For simplicity, we consider the case $N_{\rm F}=N_{\rm R}$.
It has been  shown that the BAR method leads to the minimum bias and
variance when computing the free energy difference via non-equilibrium
work methods~\cite{shirts-pande2005}.

The efficiency of the BAR method crucially depends on the switching time
$t_{\rm s}$ used to transform the initial state to the final state.  A large
switching time will make the method inefficient, while an extremely short
$t_{\rm s}$ will lead to a bias in the free energy estimate. The optimal
switching time is not known beforehand. Hence, we will use both the
JE and BAR methods with the same parametrization as in the TI scheme
and with three different switching times ($t_{\rm s}=5$, $50$ and $500$
in reduced units) to compute the interfacial free energies and compare
our results to the TI method.

Corresponding to the first step in the TI, the forward switching process
transforms the bulk LJ system into that in contact with impenetrable
flat walls in the $z$ direction.  For this step the reverse process
consists of gradually switching off of the flat walls in contact with
the system to revert back to the bulk liquid or crystal with periodic
boundary condition in all directions. Similarly, corresponding to the
second TI step, the forward switching process consists of a LJ system
in contact with a flat wall being transformed to one in contact with
the structured wall. In the reverse process, the structured walls are
gradually switched off and the flat walls switched on, such that
a liquid or crystal in contact only with a flat wall is obtained as the
final state.

\subsection{$\gamma$ from Gibbs-Cahn integration}
While the non-equilibrium work method might be computationally
advantageous compared to the TI scheme, it is still inefficient if one
needs to compute the interfacial free energies at many points along
a coexistence curve. Carrying out simulations at each point would be
cumbersome and require a huge computational effort. In such cases it
is computationally advantageous to use the ``Gibbs-Cahn integration''
method, where a differential equation is solved to obtain  interfacial
free energies along a coexistence curve if the value at one point on
the curve is known beforehand from the TI or BAR method.

The Gibbs-Cahn integration method has been used to obtain the
interfacial free energy of a hard-sphere fluid in contact with
a flat wall~\cite{laird10}, the crystal-liquid interfacial free
energy for LJ systems~\cite{laird09}, and the crystal-vapor
and crystal-liquid interfacial free energies for metallic
systems~\cite{frolov-mishin09a,frolov-mishin09b}.  In this work,
Gibbs-Cahn integration is used to compute, for a flat wall, the
wall-liquid free energy $\gamma_{\rm wl}$ along coexistence as well as
along an isotherm varying the density of the liquid.

The Gibbs-Cahn differential equation for $\gamma_{\rm wl}$ along the
pressure-temperature ($P-T$) coexistence curve is given by \cite{laird10}
\begin{equation}
d(\gamma /T)=- \frac{e_{N}+Pv_{N}}{T^2}dT + \frac{v_{N}}{T} dP ,
\label{Eq:GC_main_eqn}
\end{equation}
with $v_{N}$ the interfacial excess volume,
\begin{equation}
v_{N} = 
\int_{0}^{\infty} 
\left[ 1 - \frac{\rho(z)}{\rho}\right] dz,
\label{eq:vn}
\end{equation}
and $e_{N}$ the interfacial excess energy,
\begin{equation}
e_{N} =
\rho e
\int_{0}^{\infty}\left(\frac{e(z)}{e}-1\right) dz .
\label{eq:en}
\end{equation}
In Eqs.~(\ref{eq:vn}) and (\ref{eq:en}), $\rho$ and $e$ are the bulk
density and the bulk energy per particle, respectively.  $\rho(z)$ and
$e(z)$ are respectively the corresponding density and energy profiles
along the $z$ direction, i.e.~the direction perpendicular to the walls.
If the value of $\gamma_{\rm wl}$ is known at any point along the
coexistence curve by the TI or BAR method, Eq.~(\ref{Eq:GC_main_eqn})
can be used to determine $\gamma_{\rm wl}$ at any other point along
coexistence.  Following the derivation of a GC integration equation for
calculating $\gamma_{\rm cl}$ in Ref.~\cite{laird09}, probably one can obtain a
similar equation for computing $\gamma_{\rm wc}$ for a crystal in contact
with a flat wall, though none exists in the literature. However, in this work 
we use the GC integration method only for calculating $\gamma_{\rm wl}$.

The Gibbs-Cahn integration method is most useful when interfacial
free energies are needed along a parameter space where an intensive
thermodynamic variable such as pressure or temperature is varied.  Since
for structured wall interfaces we compute the interfacial free energies as
a function of the wall density and interaction strength between the bulk
phase and wall, the GC method is not suitable for such cases. Therefore,
we use only the TI and BAR methods for computing the interfacial free
energies for liquid or crystal in contact with a structured wall.

\subsection{Simulations}
\label{sec_sim}
Molecular dynamics (MD) computer simulations are performed at constant
particle number $N$, constant volume $V$, and constant temperature $T$.
To keep the temperature constant, the velocities of the particles were
drawn from a Maxwell-Boltzmann distribution at the desired temperature
every $200$ time steps.  The density $\rho=N/V$ of the crystal and
liquid at coexistence, corresponding to the modified LJ potential 
considered in this work, were taken from Ref.~\cite{laird09}.  As in our
earlier work, only the (111) orientation of the LJ crystal was considered.
To integrate the equations of motion, the velocity form of the Verlet
algorithm was used with a time step of $\Delta t=0.005 \tau$ (with
$\tau=\sqrt{m\sigma^{2}/\epsilon}$).

Since the simulations were performed in the $NVT$ ensemble, it is expected
that the bulk density would differ from the given density $\rho=N/V$ at
which the simulations are carried out when walls are introduced, thereby
changing the pressure. To account for this, simulations were done at
three to four different system sizes with respect to a variation of the
distance between the walls, $L_{\rm z}$, in $z$-direction, keeping the lateral
dimensions ($L_{\rm x} \times L_{\rm y}$) constant.  Interfacial free energies
were obtained by extrapolating the results to $1/L_{\rm z}=0$.  To reduce
finite size effects, lateral sizes of about $12-16\sigma$ were considered
and along the $z$ direction the length of the simulation cell varied from
about $30\sigma$ for the smallest system sizes to around $90\sigma$ for
the largest system sizes.  The smallest value of $L_{\rm z}$ considered was
large enough to avoid any spurious effects due to interactions between the
two wall-liquid (or crystal) interfaces. For the considered systems, the 
total number of particles varied from around $N=7000$ to about $N=18000$.

For the TI and BAR methods, periodic boundary conditions are employed
along the $x$, $y$ and $z$ directions when the LJ system is in contact
with flat walls since we transform a bulk LJ system periodic in all
directions to a system in contact with impenetrable flat walls. In
the second step, when the flat walls were transformed to structured
walls, periodic boundary conditions are only used along the $x$ and $y$
directions.  The GC and PA simulations are  carried out with periodic
boundary conditions only along the $x$ and $y$ directions.

For the simulations involving a flat wall, $\gamma_{\text{wl}}$ was
computed along an isotherm (at the temperature $T=1.0$) at several
densities upto the coexistence density of the liquid, $0.9288\sigma^{-3}$,
using PA, TI, BAR and the GC integration methods.  $\gamma_{\text{wl}}$
was also computed as a function of temperature along the melting
curve at five different temperatures, viz.~$T=0.618$, $0.809$, $1.0$,
$1.25$, and $1.5$ [note that $T=0.618$ corresponds to the triple point
temperature for the modified LJ potential, as given by Eq.~(\ref{eq:ur})].
At the latter temperatures, the coexistence densities of the crystal
are $0.9445\sigma^{-3}$, $0.9744\sigma^{-3}$, $1.0044\sigma^{-3}$,
$1.0411\sigma^{-3}$ and, $1.0743\sigma^{-3}$ from the lowest to the
highest temperature, respectively. The corresponding coexistence densities
for the liquid phase are $0.8282\sigma^{-3}$, $0.8829\sigma^{-3}$,
$0.9288\sigma^{-3}$, $0.9644\sigma^{-3}$, and $1.0030\sigma^{-3}$,
respectively. These coexistence densities were taken from the work of
Laird {\it et al.}~\cite{laird09}, who determined the crystal-liquid
interfacial free energy for the same potential using the GC~\cite{laird09}
and the ``cleaving potential'' approach~\cite{davidchack-laird03}. The
interfacial free energy $\gamma_{\text{wc}}$ was computed at the indicated
five different temperatures along the coexistence curve using only the
BAR and TI methods.

The statistical error bars for the TI method were calculated from $5$
different sets of data with each data set representing an average over
$10000$ configurations.  For the non-equilibrium work simulations, the
free energy difference was obtained from $3$ or $4$ different data sets
with each data set representing an average over $3$ or $4$ independent
trajectories. In each of the figures the error bars represent one standard
deviation away from the mean.

The data for GC and PA were calculated from $3-4$ independent
realizations, with each realization representing an average over two
million configurations.  While averaging over such a large number of
independent configurations is not needed for obtaining the interfacial
excess quantities used in the GC integration method, it is necessary
for computing the pressure profiles with high precision when computing
interfacial free energies via the PA method as the interfacial free
energy being the difference between the normal and tangential pressure
profiles is very sensitive to small numerical errors in the pressure
profiles~\cite{deb11}. For the GC and PA data, the error bars for
the interfacial excess quantities ($e_{\rm N}$ and $v_{\rm N}$) as well as
$\gamma_{\text{wl}}$ correspond to one standard deviation away from
the mean.

As is clear from Eq.~(\ref{Eq:GC_main_eqn}), the bulk pressures must
also be calculated at each temperature or density to compute the
integrals. Since we carry out simulations in the $NVT$ ensemble,
the bulk pressures would be different for different system sizes
along the $z$ direction. Hence, we calculated the bulk pressures,
performed the integration at each system size, and then extrapolated
the results to $1/L_{\rm z}=0$.  To compute the interfacial excess energy
and the interfacial excess volume for use in the GC integration method,
a bulk region must be identified in the middle of the simulation cell
and away from the walls. We choose a length of $5\sigma$ in the middle
of the simulation cell to calculate the bulk density, pressure and
energy per particle and the same length was considered for all system
sizes. To calculate the density, pressure and energy profiles a bin
size of $0.05\sigma$ was chosen along the $z$ direction. All integrals
in the TI method as well as the GC integration method were performed
by the trapezoidal rule and it was checked that numerical integration
errors were less than the statistical errors.

\section{Results}
\label{sect:results}
For a  direct comparison of the TI, PA, GC and BAR methods, we computed
$\gamma_{\rm wl}$ for the case of a liquid in contact with a flat wall.
Since the PA method and probably also the GC approach cannot be used
to compute $\gamma_{\rm wc}$, this quantity is only determined via the
TI and BAR methods. In the following, different wetting conditions
are mainly studied for one state at coexistence, corresponding to the
temperature $T=1.0$ and the crystal and liquid coexistence densities at
$1.0044\,\sigma^{-3}$ and $0.9288\,\sigma^{-3}$, respectively. For this case,
$\gamma_{\rm wl}$ and $\gamma_{\rm wc}$ are investigated as a function
of the interaction parameters $\epsilon_{\rm sw}$ and $\rho_{\rm
sw}$. Note that for the structured wall only the TI and BAR methods are
employed. We first present the results corresponding to the interfaces
with the flat wall and then discuss the case of the structured wall.

\subsection{Flat wall}
\begin{figure}
\includegraphics[width=3.0in]{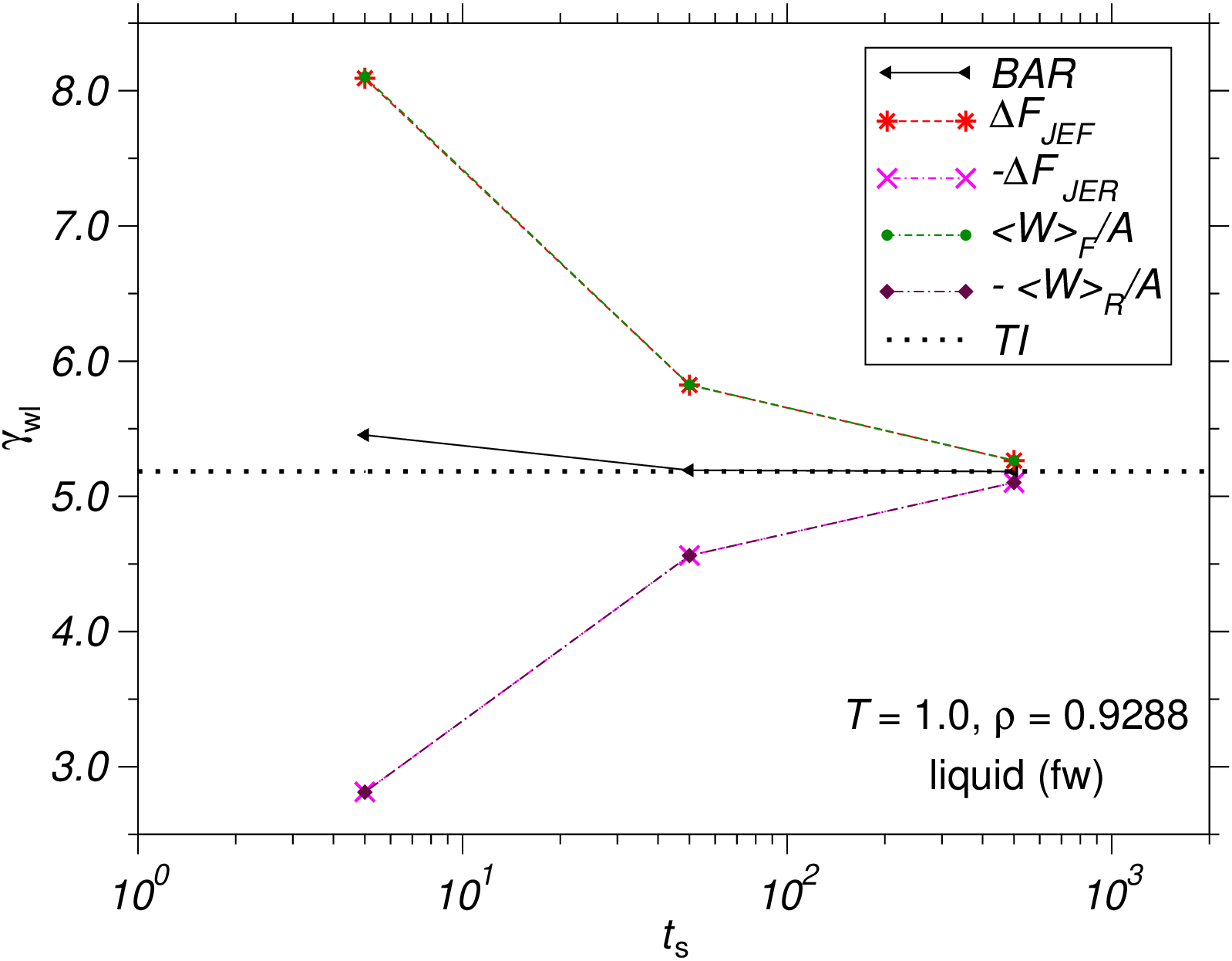}
\caption{\label{fig:gam_new_wl}(Color online)
Estimates of $\gamma_{\text{wl}}$ for the flat wall (fw) at the
indicated values of $T$ and $\rho$ from non-equilibrium work methods
(see text). The horizontal dotted line is the result from TI. All
results are obtained from an extrapolation to $L_{\rm z}^{-1}=0$. Error
bars are not shown as they are less than the symbol size.}
\end{figure}
\begin{figure}
\includegraphics[width=3.0in]{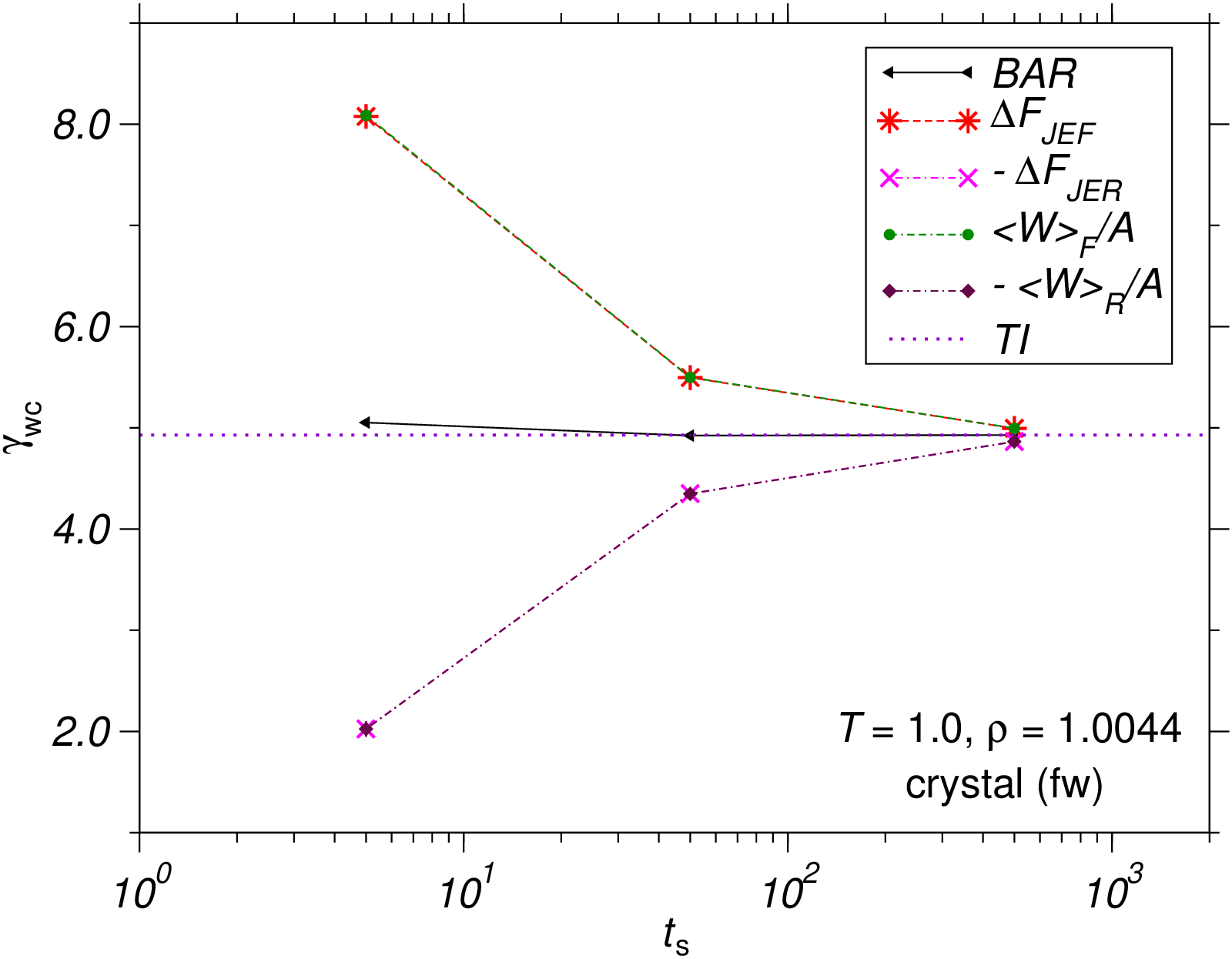}
\caption{\label{fig:gam_new_wc}(Color online)
$\gamma_{\text{wc}}$ vs.~$t_{\rm s}$ at the indicated values of $T$
and $\rho$ obtained using TI and several non-equilibrium work methods. 
Error bars are less than the symbol size.}
\end{figure}

We first show results obtained from the four different methods at
temperature $T=1.0$ and the coexistence densities $0.9288\,\sigma^{-3}$
and $1.0044\,\sigma^{-3}$ for the liquid and the crystal, respectively.
In the following, we denote the estimated interfacial free energy from JE
[cf.~Eq.~(\ref{eq:jarest})] for the forward switching process by $\Delta
F_{\rm JEF}$ and that for the negative of the reverse switching process
by $-\Delta F_{\rm JER}$. We compare $\Delta F_{\rm JEF}$ and $-\Delta
F_{\rm JER}$ to the correspondent average work values (divided by the
interface area $A$), $\langle W \rangle_{\rm F}/A$ and $-\langle W
\rangle_{\rm R}/A$, respectively. Estimates of $\gamma_{\text{wl}}$ and
$\gamma_{\text{wc}}$ from the non-equilibrium work methods as a function
of switching time $t_{\rm s}$ are shown in Figs.~\ref{fig:gam_new_wl}
and \ref{fig:gam_new_wc}, respectively. Note that all the results in
these figures are obtained from an extrapolation to $L_{z}^{-1}=0$
(see above). Since, $\langle W \rangle_{\rm F} > \Delta F$, it follows
that $\langle W \rangle_{\rm R} > -\Delta F$ or $-\langle W \rangle_{\rm R} <
\Delta F$. These relations are rationalized by our calculations (see
figures).

Figures~\ref{fig:gam_new_wl} and \ref{fig:gam_new_wc} show that, as the
switching time $t_{\rm s}$ increases, $\Delta F_{\rm JEF}$ and $-\Delta
F_{\rm JER}$ gradually converge to the interfacial free energies, as
obtained from the TI scheme (horizontal dotted lines). As indicated in
the figures, the JE values are similar to the corresponding average work
values. Whereas at small switching times ($t_{\rm s} \lesssim 50$) the
JE estimates significantly differ from the TI value, for $t_{\rm s}=500$
the deviation between the JE and TI values is less than 2\% (this also
holds for the corresponding average work values). On the other hand,
the free energy estimated by the BAR method is in excellent agreement
with the TI value already at $t_{\rm s}=50$, with difference between
the two estimates being less than 0.3\%.  However, at $t_{\rm s}=50$,
the error bars corresponding to the BAR method are larger than those of
the TI values. Only at $t_{\rm s}=500$, the error bars for the BAR and
TI method have the same order of magnitude. Similarly, the magnitude of
the error bars corresponding to the JE estimates and the average work
values reduce to those of the TI method at $t_{\rm s}=500$.

We find that the BAR method (with $t_{\rm s}=50$) is twice as fast as
the TI approach, though, for $t_{\rm s}=50$, the error bars for the
BAR method are slightly larger.  At $t_{\rm s}=500$, the error bars and
also the computational load for the BAR and the TI method are similar
and the BAR method is no more computationally advantageous compared to
the TI approach.  On the other hand, the computational load needed for
the JE estimates and the corresponding average work values, at $t_{\rm
s}=500$, is about one third that of the TI method, which is still faster
compared to the BAR method at the smaller switching time $t_{\rm s}=50$.
Since the BAR approach requires information from both the forward and
reverse trajectories its efficiency is reduced as compared to the JE
estimates, which need calculations only in one direction, either forward
or reverse.

The good agreement of the average work values for $t_{\rm s}=500$ with the
TI results, shows that at this switching time the switching process is
quasi-static and hence is akin to a TI.  In our TI simulations, carried
out independently of the non-equilibrium work simulations, we choose a
much larger equilibration time before the production runs thus increasing
the computational load.  Hence, an advantage of carrying out simulations
with various switching times to calculate the non-equilibrium work,
allows us to identify the switching time required for the transformation
from the initial to the final state to be quasi-static and reduce to a
conventional TI.  The good agreement of the average work values with JE
estimates at $t_{\rm s}=500$ and with the TI data also indicates that
there is no advantage in using the non-equilibrium work method over a TI
approach as regards computational speed is concerned. However, carrying
out simulations for both forward and reverse switching processes help
us to check for any hysteresis in the TI. The convergence of both the
forward and reverse work values (divided by $A$) to the interfacial
free energy obtained from TI, indicates the lack of hysteresis in our
TI scheme. Secondly, the good agreement between the non-equilibrium work
and TI method also validates the accuracy of our TI method.

In the remainder of this subsection, unless otherwise indicated, we show
results pertaining to the non-equilibrium work approach obtained using
only the BAR method for the switching time $t_{\rm s}=500$.

To determine the interfacial free energy along an isotherm at different
densities or along a coexistence curve at different temperatures via the
GC integration we need as input $\gamma_{\rm wl}$ at one point along the
curve as well as the excess volume (in case of determining $\gamma_{\rm
wl}$ along the coexistence curve, also the excess energy) at all points
along the curve at which $\gamma_{\rm wl}$ is to be determined.

\begin{figure}
\includegraphics[width=3.0in]{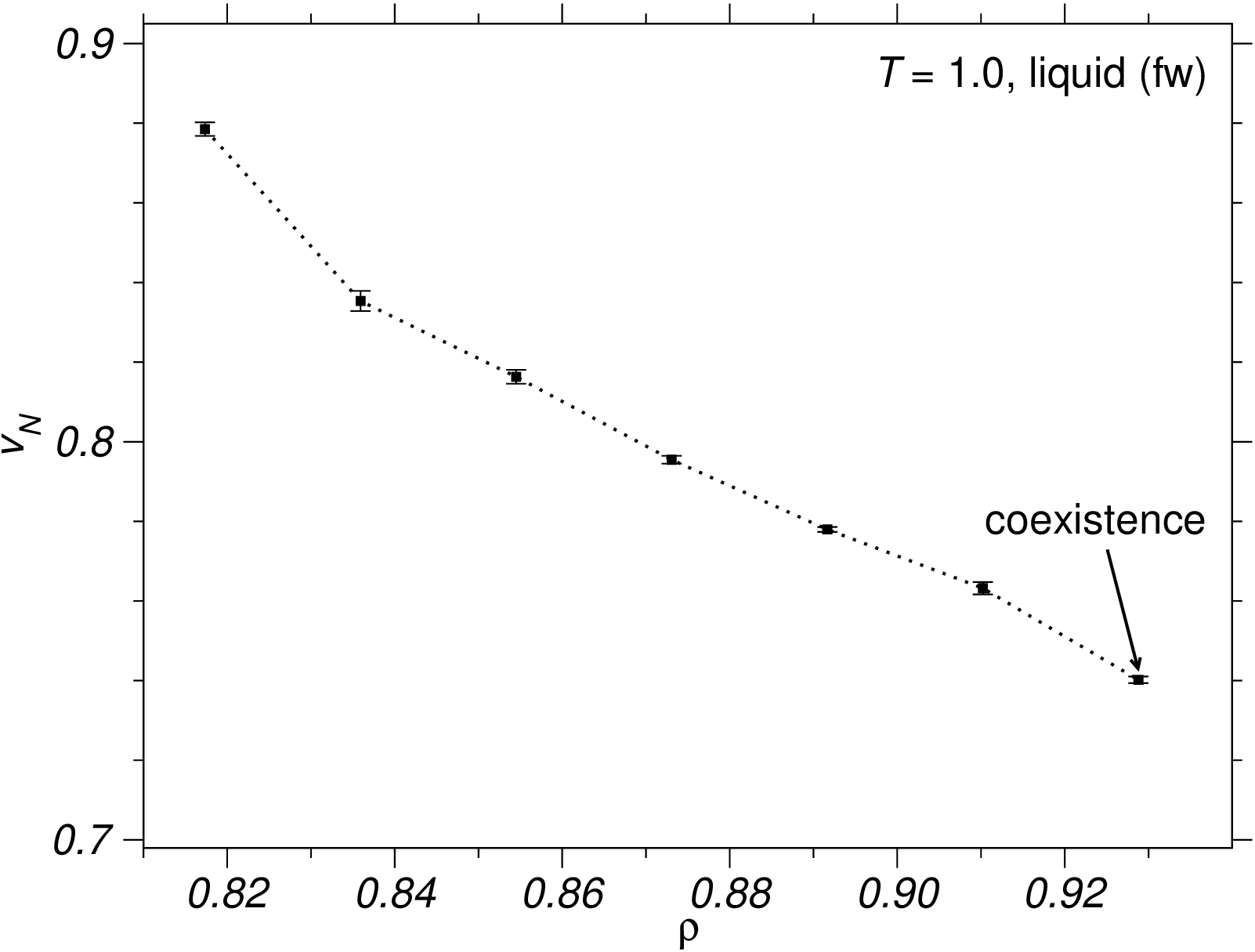}
\caption{\label{fig:excess_vs_rho}(Color online)
Excess interfacial volume $v_{\rm N}$ as a function of density at $T=1.0$,
for the liquid in contact with a flat wall (fw).}
\end{figure}
\begin{figure}
\includegraphics[width=3.0in]{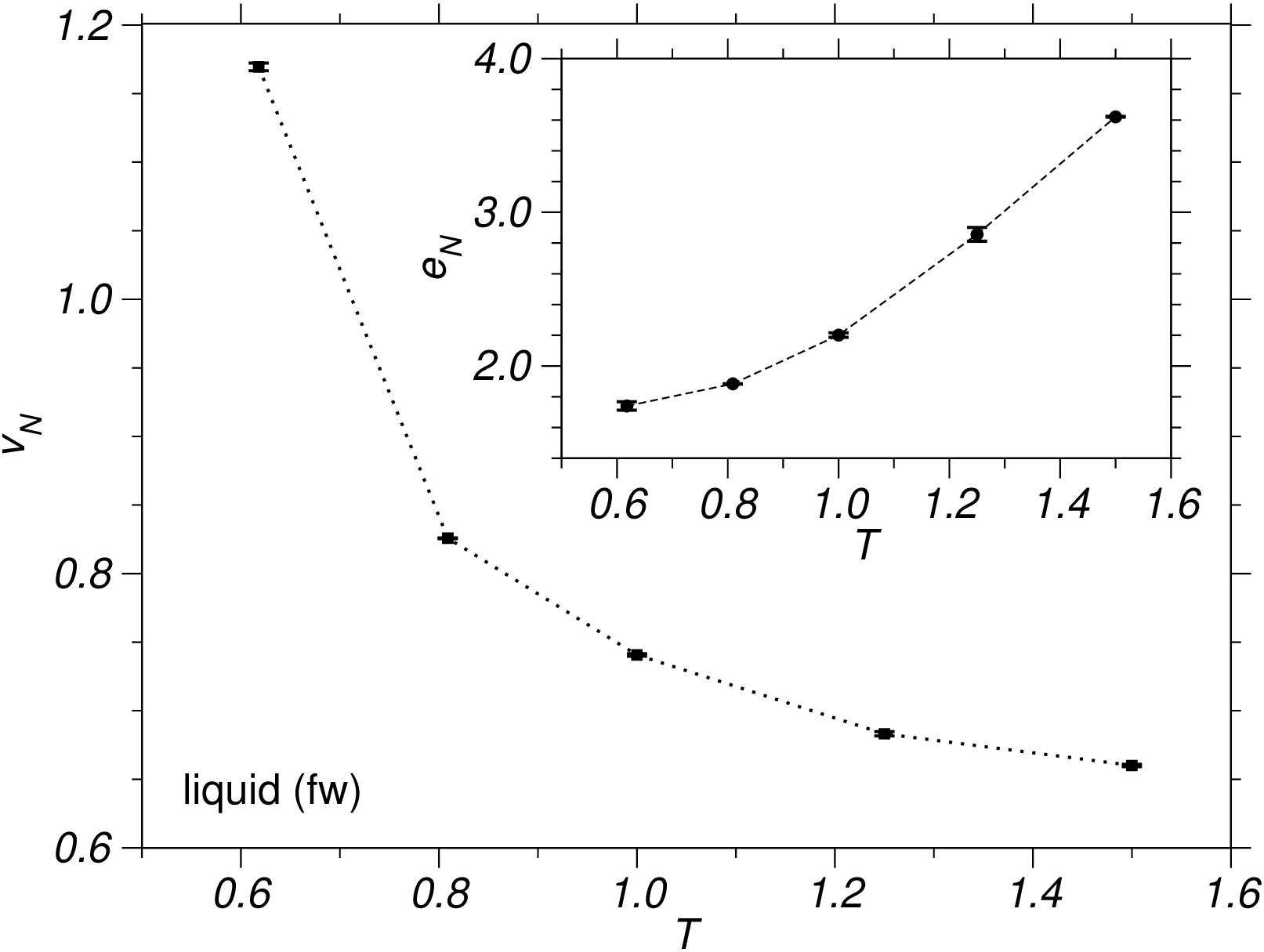}
\caption{\label{fig:excess_vs_T}(Color online)
Excess interfacial volume $v_{\rm N}$ and (in the inset) excess interfacial
energy ($e_{\rm N}$) as a function of the temperature along the liquid-crystal
coexistence curve for liquid in contact with a flat wall.}
\end{figure}
\begin{figure*}
\begin{center}$
\begin{array}{cc}
\includegraphics[width=3.0in]{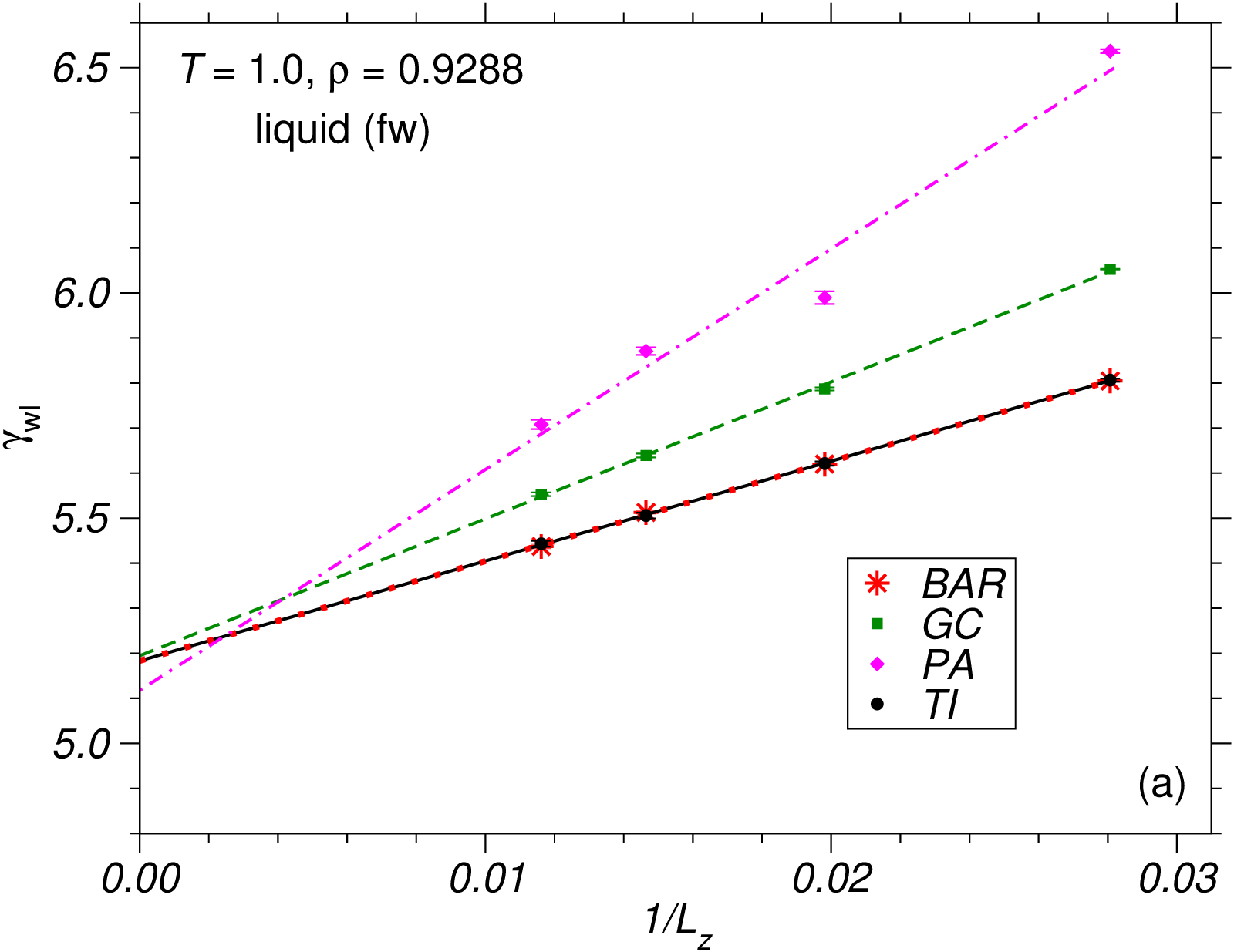} &
\includegraphics[width=3.0in]{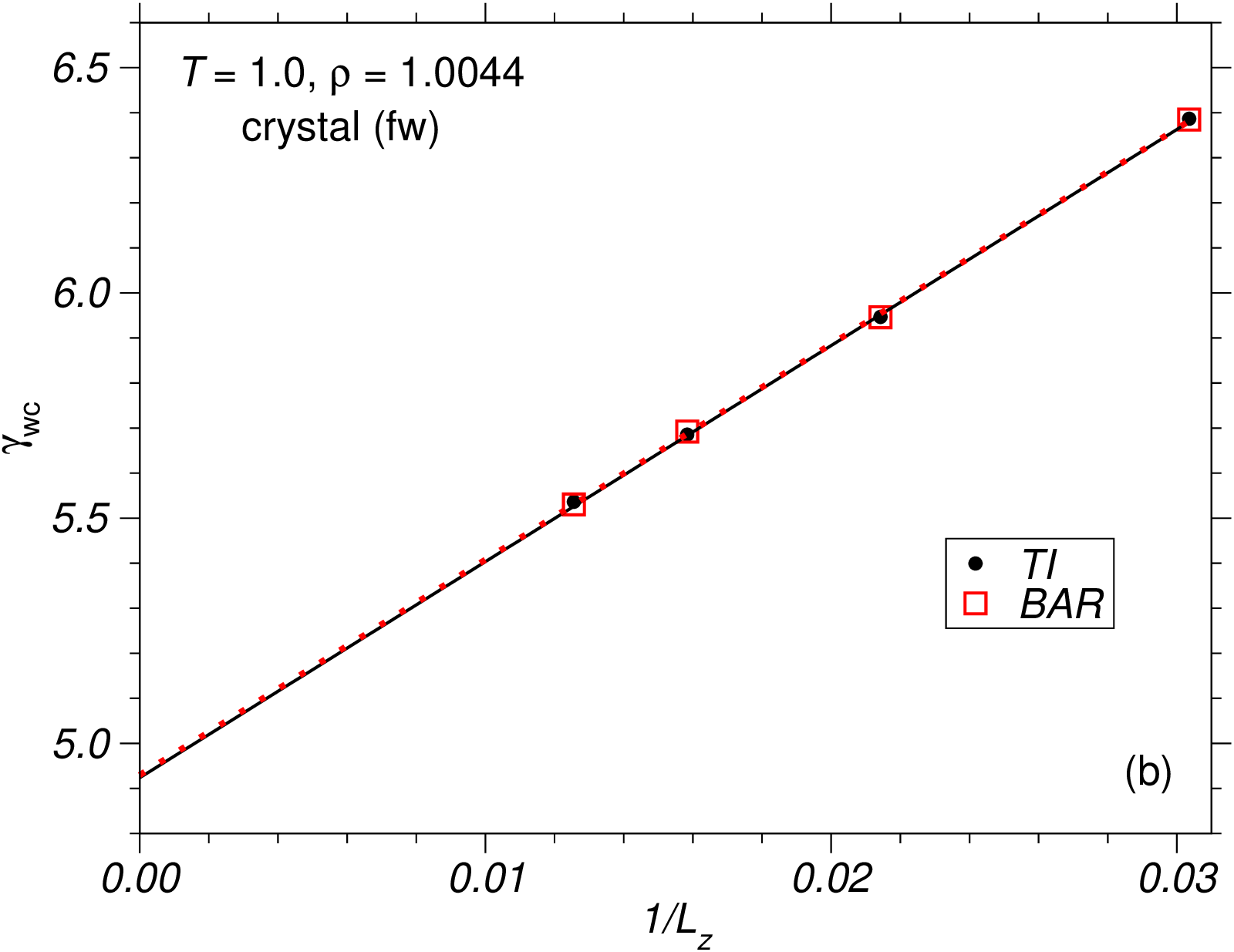}
\end{array}$
\end{center}
\caption{\label{fig:ife_fw_l_c_Lz}(Color online)
Interfacial free energies of the (a) liquid in contact with the flat
wall and (b)(111) orientation of the crystal in contact with the flat wall
for different system sizes $L_{\rm z}$ of the simulation cell along the $z$
direction at the coexistence temperature $T=1.0$. The coexistence densities of the
liquid and crystal at this temperature are $0.9288\sigma^{-3}$ and $1.0044\sigma^{-3}$
respectively. For the wall-liquid
interfaces, the interfacial free energy obtained from the PA, GC,
TI and BAR methods are shown, while for the wall-crystal simulations,
data obtained from the TI and BAR methods are plotted. The straight
lines connecting the points denote a linear least-squares fit from
whose intercepts the interfacial free energies are determined. In (b)
error bars are less than the symbol size.}
\end{figure*}
\begin{figure*}
\begin{center}$
\begin{array}{cc}
\includegraphics[width=3.0in]{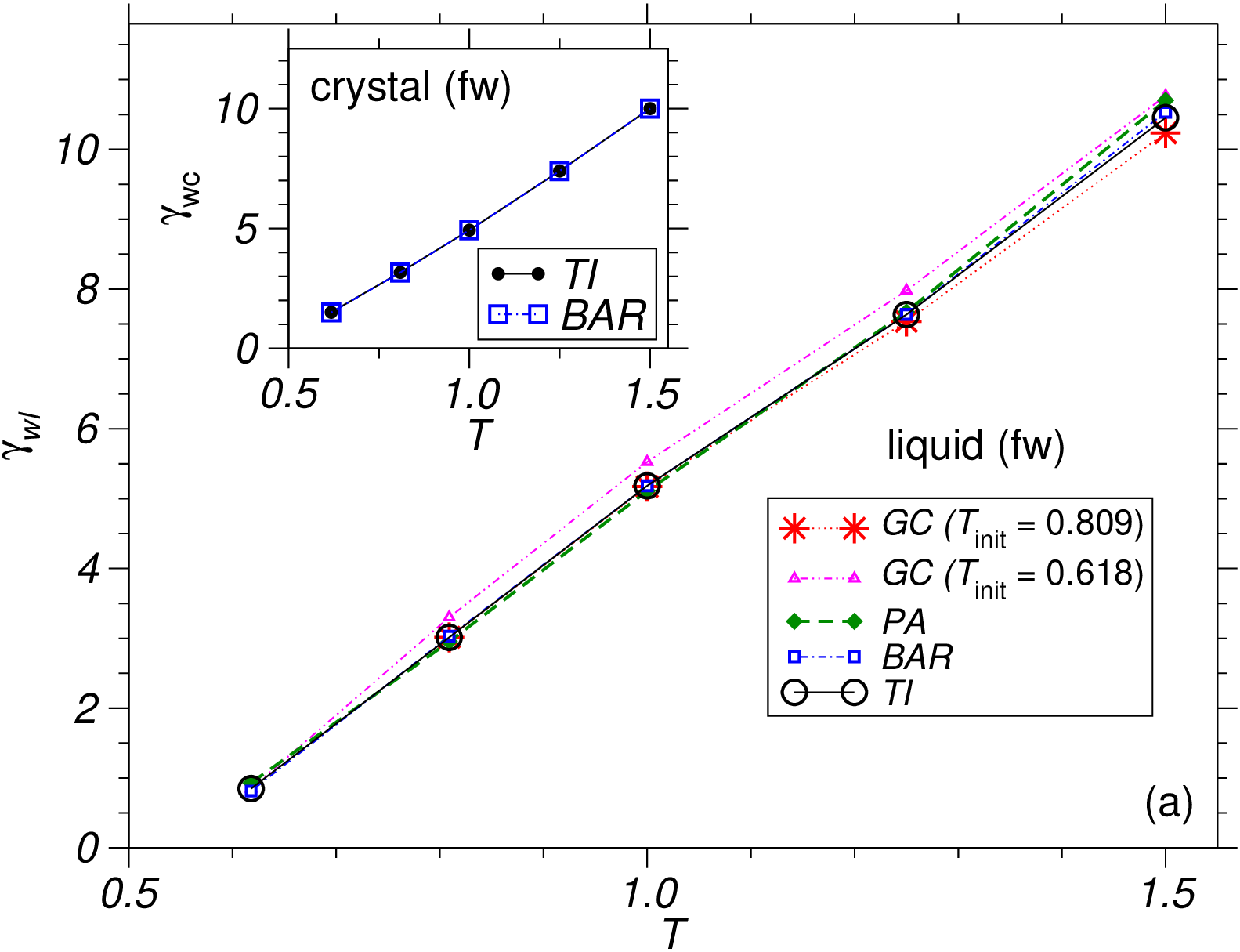} &
\includegraphics[width=3.0in]{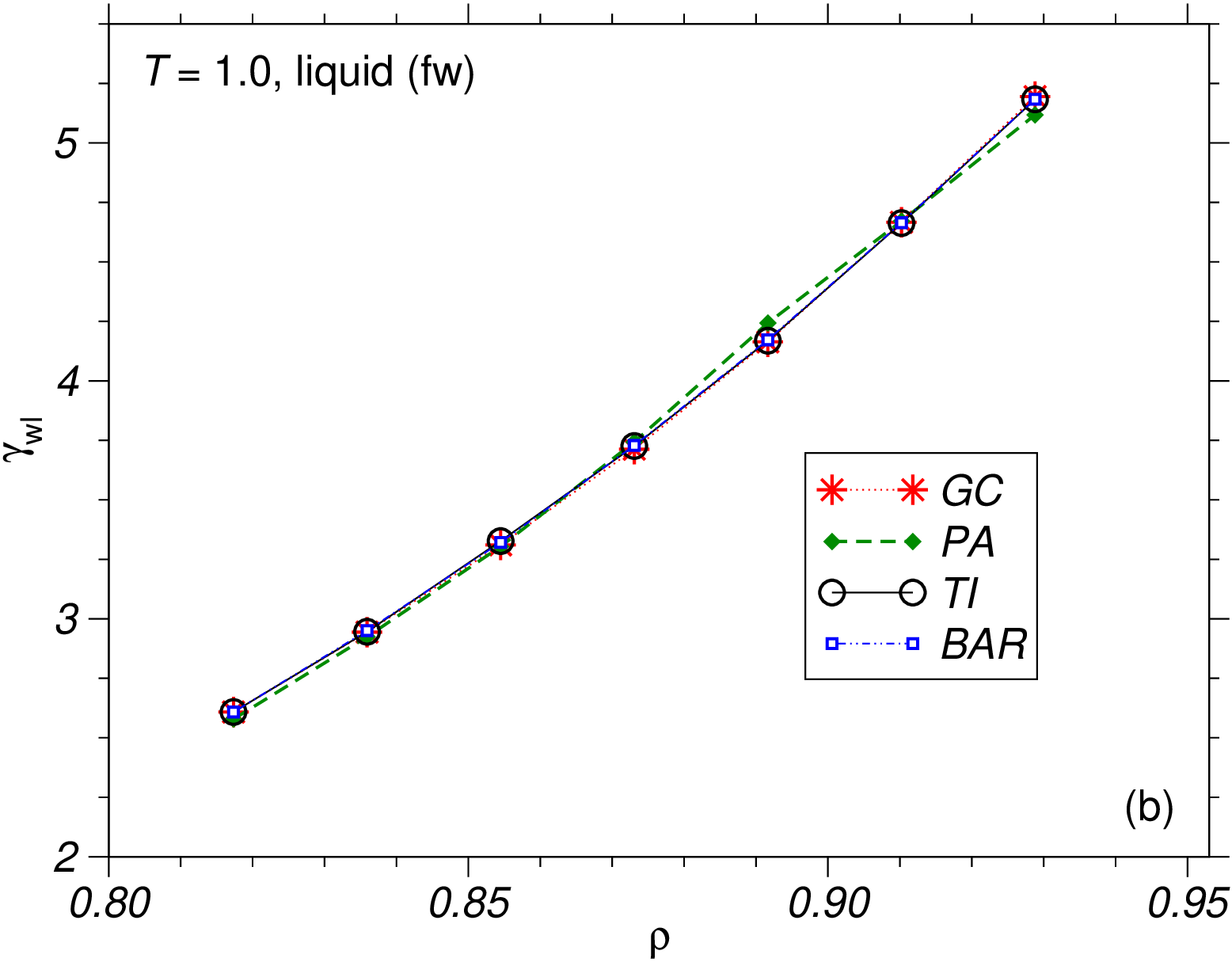}
\end{array}$
\end{center}
\caption{\label{fig:ife_fw_T_P}(Color online)
(a) $\gamma_{\text{wl}}$ and $\gamma_{\text{wc}}$ (shown in the
inset) as a function of the melting temperature and (b) $\gamma_{\rm wl}$
vs. density of the liquid  at $T=1.0$.  In (a) two sets of GC 
data are shown for $\gamma_{\text{wl}}$. One corresponds to data where
the initial known value of the interfacial free energy is taken from
the TI data at $T=0.618$ and the values for $T>0.618$ obtained via GC
integration. The second curve corresponds to the case where the known
value of $\gamma_{\text{wl}}$ was taken from the TI data at $T=0.809$
and for $T>0.809$, the interfacial free energies are obtained using GC
integration. In case of (b), the known value of $\gamma_{\text{wl}}$
was taken from the TI data at the lowest value of the given liquid
density. The data corresponding to the BAR method represent results for
a switching time $t_{\rm s}=500.0$. Error bars in (a) and (b) are less than
the symbol size.}
\end{figure*}
Figure~\ref{fig:excess_vs_rho} shows the excess volume as a function of
density up to the coexistence density at $T=1.0$. Since the temperature
is constant ($dT=0$), the first term in Eq.~(\ref{Eq:GC_main_eqn})
is zero and we only need the interfacial excess volume, $v_N(\rho)$
and one reference value for $\gamma_{\rm wl}$.  Here, this reference
value was taken from the TI result corresponding to the lowest density
at $\rho_{\rm l}=0.811\sigma^{-3}$.  Note that the lateral system sizes
were kept at $L_{x}=13.5892\sigma$ and $L_{y}=15.691\sigma$ for the
different densities.  From the lowest to the highest density, the system
sizes along the $z$ direction were $66.907\,\sigma$, $65.420\,\sigma$,
$63.998\,\sigma$, $62.636\,\sigma$, $61.331\,\sigma$, $69.678\,\sigma$,
and $68.285\,\sigma$, respectively.  The number of particles for all
densities up to $\rho_{\rm l}=0.8916\sigma^{-3}$ was $N=11661$. For
$\rho_{\rm l}=0.9102\,\sigma^{-3}$ and $\rho_{\rm l}=0.9288\,\sigma^{-3}$,
$N=13524$ was chosen.

The interfacial excess quantities were also computed for all the
other system sizes from which the final value of $\gamma_{\rm wl}$
was determined by an extrapolation to $L_{z}^{-1}=0$.  At all other
system sizes, the interfacial excess quantities had similar values and
varied in the same manner along the isotherm at the different densities.
There was no clear dependence of the interfacial excess quantities on
the system size.  As the density increases and approaches the liquid
coexistence density at this temperature, $0.9288\,\sigma^{-3}$, the
bulk pressures in the liquid also increase.  Similar to the hard-sphere
case~\cite{laird09}, we also find that the excess volume decreases with
increasing density (or bulk pressure).

\begin{figure}
\includegraphics[width=3.0in]{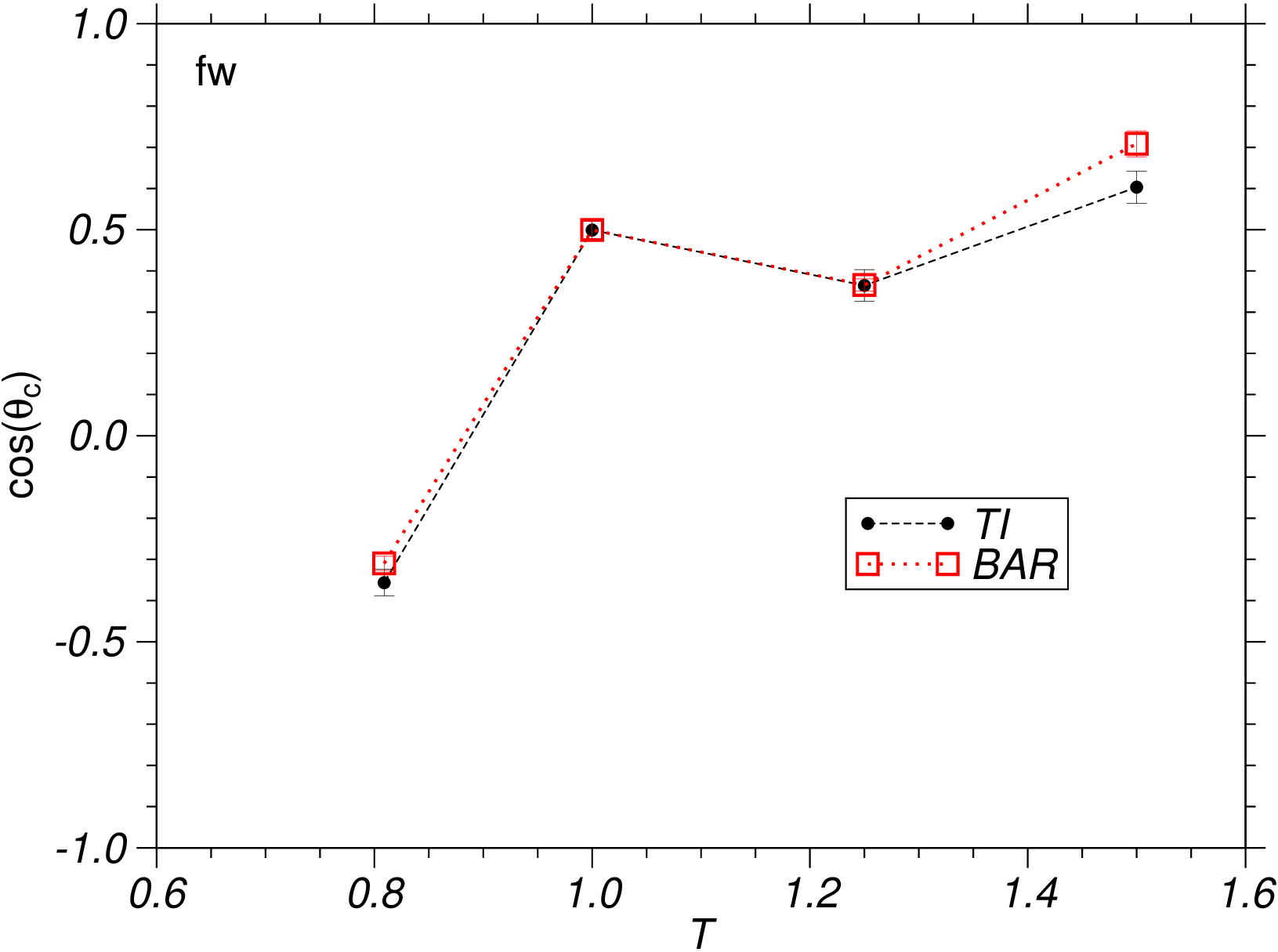}
\caption{\label{fig:contang_fw}(Color online)
Cosine of the contact angle as a function of temperature along
liquid-crystal coexistence for system in contact with a flat wall, obtained 
from Young's equation by using $\gamma_{\rm wl}$ and $\gamma_{\rm wc}$ from 
the TI and BAR data of Fig.~6a and using values of $\gamma_{\text{cl}}$ as 
reported in Ref.~\cite{davidchack-laird03}. Error bars are less than the 
symbol size.}
\end{figure}
Figure~\ref{fig:excess_vs_T} shows the excess volume and excess energy as
function of temperature at coexistence.  The system sizes $L_{x}\times
L_{y}\times L_{z}$ corresponding to the different temperatures are
as follows: $13.871\times 16.016 \times 63.378$ at $T=0.618$, $13.727
\times 15.851 \times 70.397$ at $T=0.809$, $13.589 \times 15.691 \times
68.285$ at $T=1.0$, $13.428 \times 15.505 \times 67.357$ at $T=1.25$,
and $13.288 \times 15.344 \times 66.134$ at $T=1.5$ (with all lengths in
units of $\sigma$). The number of particles was $N=11661$ at $T=0.618$ and
$N=13524$ at the other temperatures. It is observed that the excess volume
decreases with temperature while the excess energy shows an increase
with temperature.  We note that the interfacial excess quantities do
not exhibit finite-size effects for the considered system sizes.

\begin{figure}
\includegraphics[width=3.0in]{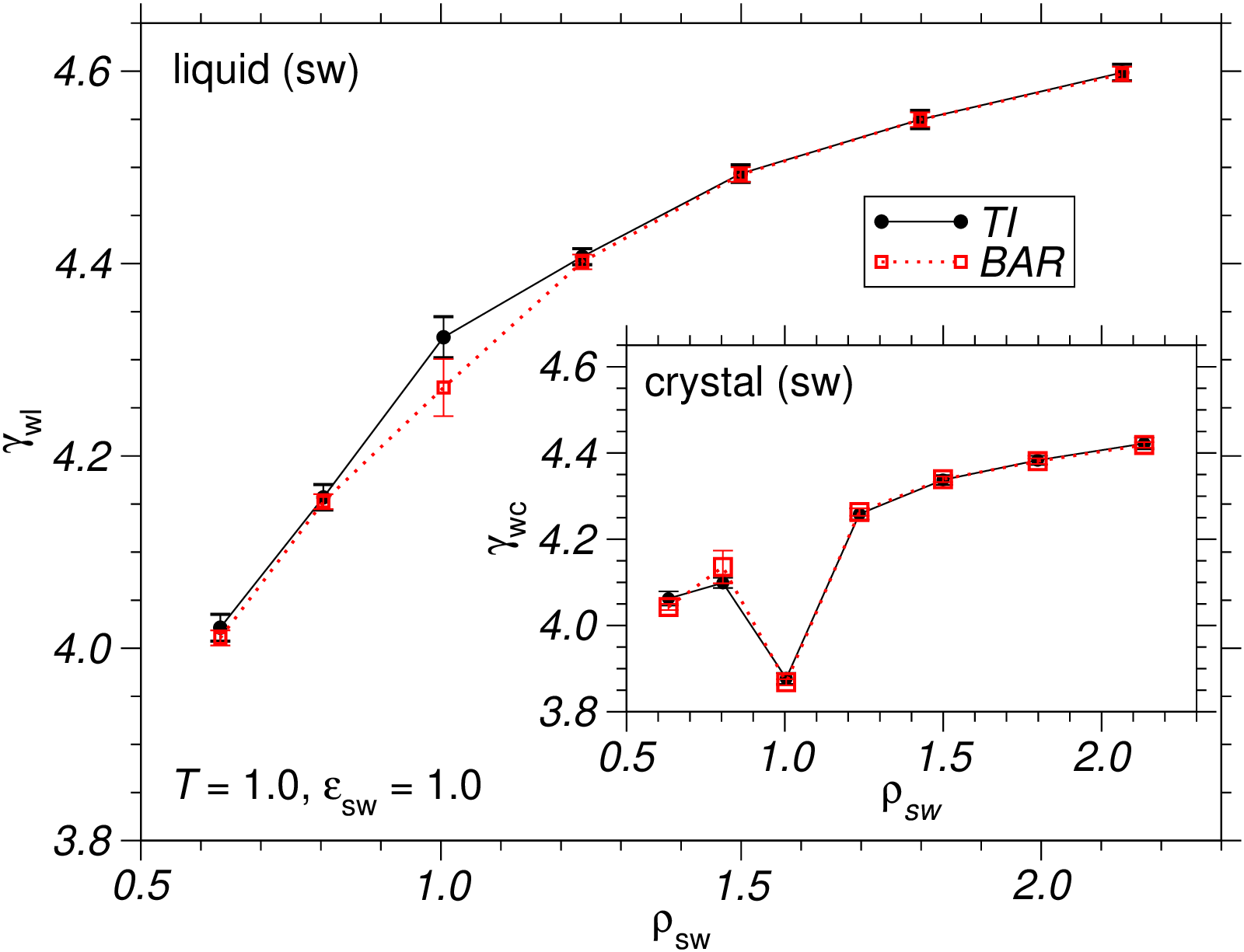}
\caption{\label{fig:ifesw_vs_rhow}(Color online)
$\gamma_{\text{wl}}$ and $\gamma_{\text{wc}}$ (shown in the inset) as
function of $\rho_{\rm sw}$ at the temperature $T=1.0$ and $\epsilon_{\rm
sw}=1.0$, obtained from the BAR (with $t_{\rm s}=500$) and the TI method.}
\end{figure}
\begin{figure}
\includegraphics[width=3.0in]{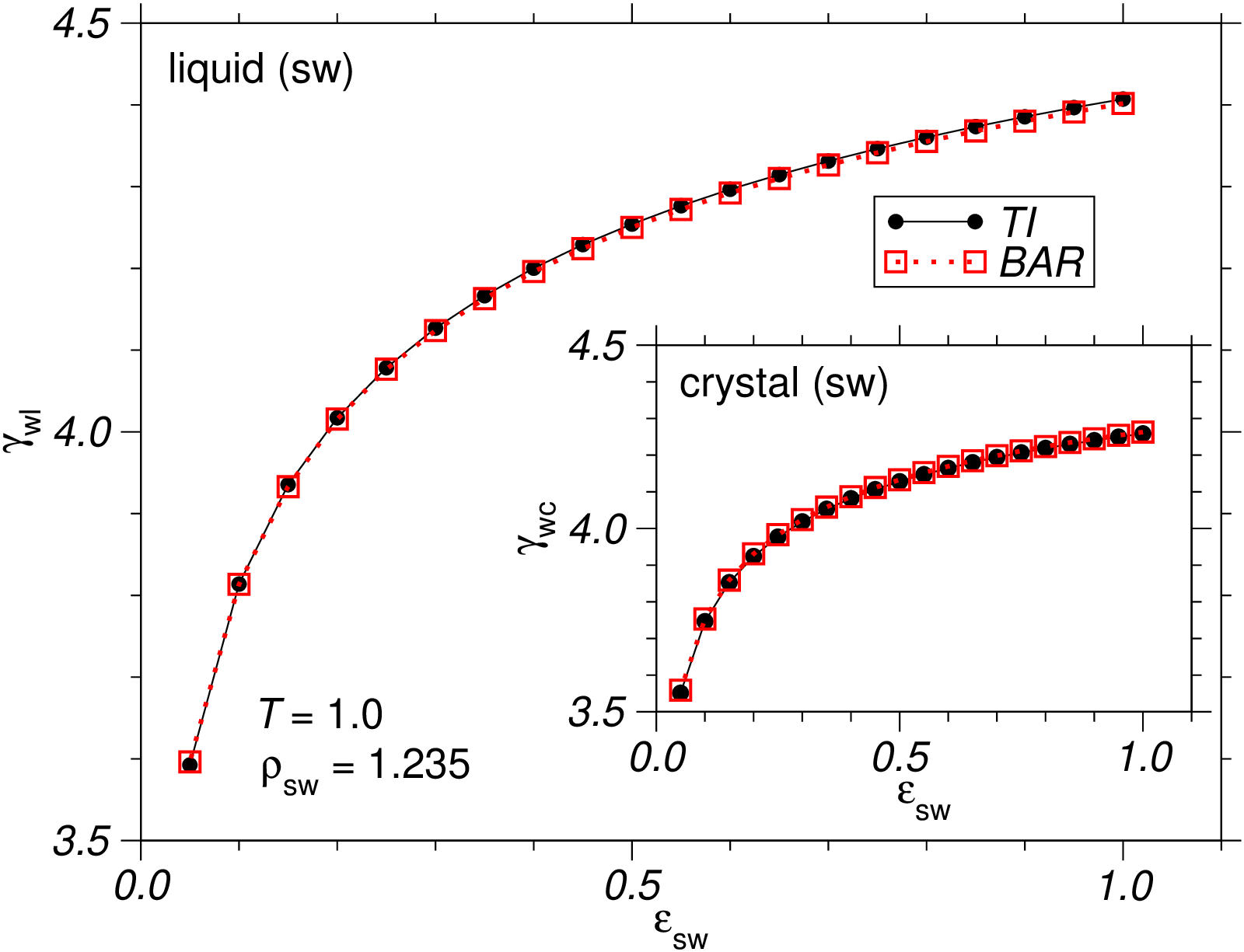}
\caption{\label{fig:ifesw_vs_ew}(Color online)
$\gamma_{\text{wl}}$ and $\gamma_{\text{wc}}$ (shown in the inset) versus
the interaction strength $\epsilon_{\rm sw}$ at the temperature $T=1.0$
and $\rho_{\rm sw}=1.235$ obtained using the BAR (with $t_{\rm s}=50$)
and TI methods. Error bars are less than symbol size.}
\end{figure}
\begin{figure}
\includegraphics[width=3.0in]{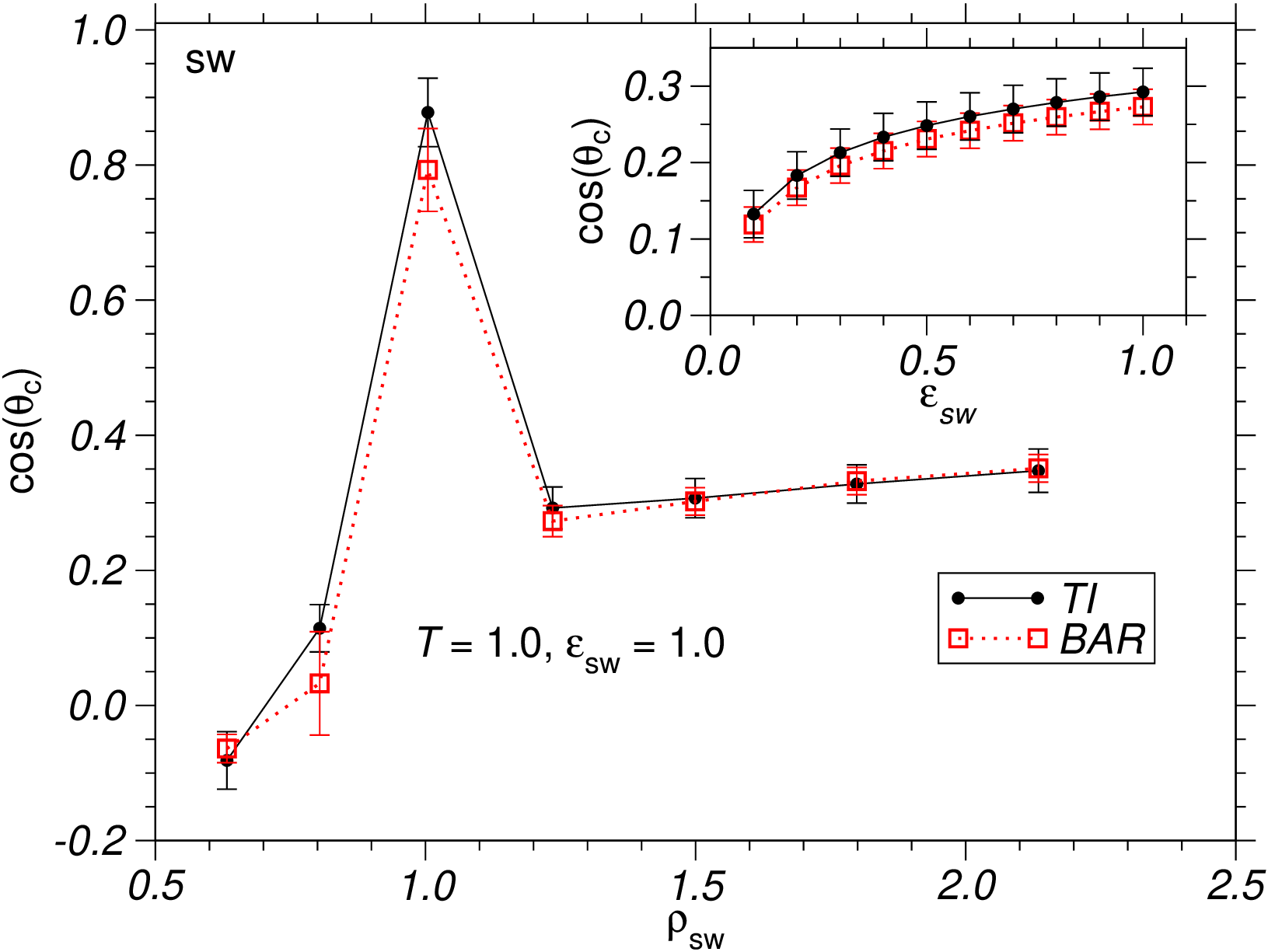}
\caption{\label{fig:contang_sw}(Color online) 
Cosine of the contact angle obtained from Young's equation at the
temperature $T=1.0$ as a function of $\rho_{\rm sw}$ at $\epsilon_{\rm
sw}=1.0$ and in the inset versus the interaction strength  $\epsilon_{\rm
sw}$ at $\rho_{\rm sw}=1.235$. $\gamma_{\rm wl}$ and $\gamma_{\rm wc}$
were taken from the BAR and TI data of Figs.~\ref{fig:ifesw_vs_rhow}
and \ref{fig:ifesw_vs_ew} and values of $\gamma_{\rm cl}$ reported in
Ref.~\cite{davidchack-laird03} were used.}
\end{figure}
For each individual run and system size, the interfacial excess
quantities and the GC integration was carried out independently
to obtain $\gamma_{\text{wl}}$ along an isotherm as a function of
density and along the melting curve as a function of temperature. From
the obtained values of $\gamma_{\text{wl}}$ for the four considered
system sizes, an extrapolation was performed to $L_{z}^{-1}=0$ and the
final values were calculated from the intercepts with the ordinate.
In Figs.~\ref{fig:ife_fw_l_c_Lz}a and \ref{fig:ife_fw_l_c_Lz}b, we show
the results for the interfacial free energy from the various methods as
a function of $1/L_{z}$ for the liquid and crystal in contact with the
flat wall, respectively.  While the extrapolated values obtained from
the BAR, TI and GC methods are in good agreement with each other up to
the numerical acccuracy, results obtained from the PA method deviate
from the rest by $2-3\%$.

Figure \ref{fig:ife_fw_T_P}a shows $\gamma_{\rm wc}$ and
$\gamma_{\text{wl}}$ as a function of temperature along the melting
curve, and in Fig.~\ref{fig:ife_fw_T_P}b, $\gamma_{\rm wl}$ is
displayed as a function of density at $T=1.0$.  For the crystal, we
find very good agreement between the BAR and the TI methods (inset of
Fig.~\ref{fig:ife_fw_T_P}).  To determine $\gamma_{\rm wl}$ along the
melting curve, the interfacial free energy obtained from the TI method at
$T=0.618$ was first chosen as the initial value for the GC integration.
With this initial value, the agreement of the GC integration data with
the TI and BAR data becomes progressively worse with temperature and
at $T=1.5$, the difference between GC and TI or BAR result is around
$9\%$. However, when the value of $\gamma_{\rm wl}$ obtained from TI
at $T=0.809$ is taken as the initial value, the agreement between the
different methods is much better. This could be attributed to the fact
that the pressure-temperature melting curve for LJ potentials that we
consider here is clearly non-linear~\cite{asano-fuchizaki12} between
$T=0.618$ and $T=0.809$ signifying that more data is needed between
these two temperatures, whereas it exhibits essentially linear behavior
for $T\geq 0.809$.

For the GC simulations to compute $\gamma_{\rm wl}$ along an isotherm
at different densities, the interfacial free energy obtained from the
TI method at the lowest density was chosen as the initial value to carry
out the integration. Figure~\ref{fig:ife_fw_T_P}b clearly shows excellent
agreement between all the four methods. The error bars corresponding to
the different methods are all of a similar magnitude.

From the values for $\gamma_{\rm wl}$ and $\gamma_{\rm wc}$
predicted by the BAR and TI methods and using the data from
Ref.~\cite{davidchack-laird03} for $\gamma_{\rm cl}$, we
determined the cosine of the contact angle as a function of
temperature along the coexistence curve via Young's equation.
The results are shown in Fig.~\ref{fig:contang_fw}. At $T=0.618$,
$(\gamma_{\text{wl}}-\gamma_{\text{wc}})/\gamma_{\text{cl}} < -1$,
corresponding to the complete drying conditions. As $T$ increases,
the flat wall prefers the crystal more than the liquid 
($\gamma_{\rm wc}<\gamma_{\rm wl}$) and incomplete
wetting is observed ($-1<\cos \theta_{\rm c}<1$). At higher temperatures
($T>1.0$) partial wetting is observed ($0<\cos \theta_{\rm c}<1$),
with $\cos \theta_{\rm c}$ varying  non-monotonically with respect to 
temperature along the melting curve. While good agreement is found in general 
between the BAR and TI methods, here, we observe a slightly larger difference 
between them as compared to that for the individual interfacial free energies. 
This is because $\cos \theta_{\rm c}$ depends on the difference between 
$\gamma_{\text{wl}}$ and $\gamma_\text{wc}$, which magnifies the relative error.

\subsection{Structured Wall}
We now consider the LJ system in contact with a structured wall, with
the interaction between the wall and the LJ system to be of the same
purely repulsive WCA potential. For the structured wall, the interesting
parameters are the interaction strength and the density of the wall. We
study the variation of the interfacial free energy as a function of
these two parameters at the liquid-crystal coexistence temperature
$T=1.0$, choosing the coexistence densities $0.9288\,\sigma^{-3}$ and
$1.0044\,\sigma^{-3}$ for the liquid and the crystal, respectively. In
the following the BAR and the TI method are compared.

In Fig.~\ref{fig:ifesw_vs_rhow}, we plot the interfacial free energies
of the LJ liquid and crystal as a function of the density of the wall
(at $\epsilon_{\rm sw}=1.0$).  In the computations with the BAR method a
switching time of $t_{\rm s}=500$ was used. The density of the structured
wall was modified by choosing different lattice constants for the fcc
structure of the wall. We observe that the interfacial free energy of
the liquid increases with increasing density of the wall.  Increasing
density of the wall means a larger repulsive force acting on the liquid
particles which enhances $\gamma_{\text{wl}}$. The general trend for
$\gamma_{\text{wc}}$ is the same. However, when the density of the wall
is the same as the equilibrium density of the crystal we find a sharp
dip in the interfacial free energy. This is expected since when the wall
has the same structure as the crystal less energy is required to create
an interface.

Figure~\ref{fig:ifesw_vs_ew} shows $\gamma_{\rm wl}$ and $\gamma_{\rm
ws}$ for different interaction strengths $\epsilon_{\rm sw}$
between the wall and the LJ system. The density of the substrate is
$\rho_{\text{sw}}=1.235$, i.e.~the density of the structured wall is
larger than that of the bulk LJ crystal. To reduce the computational
effort, $\epsilon_{\rm sw}$ was itself used as the switching parameter,
with the system at $\epsilon_{\rm sw}=1$ considered as the reference
state.  For the BAR calculations, a switching time $t_{\rm s}=50$ was
sufficient to obtain good agreement with the TI method as regards
the mean value and also the standard deviation.  All the data
corresponding to the BAR calculations in Fig.~\ref{fig:ifesw_vs_ew}
are shown for a switching time $t_{\rm s}=50$.  Increasing the interaction 
strength between the wall and the LJ system
leads to a greater repulsive interaction between them, thereby enhancing
the interfacial free energies.

As can be inferred from Figs.~\ref{fig:ifesw_vs_rhow} and
\ref{fig:ifesw_vs_ew}, at the considered thermodynamic states
$\gamma_{\rm wc}<\gamma_{\rm wl}$ holds. This signifies the
occurrence of incomplete wetting.  To determine the contact angle
from Young's equation, we use $\gamma_{\rm cl}$, as obtained by
the cleaving potential method~\cite{laird10}, and $\gamma_{\rm
wl}$ and $\gamma_{\rm wc}$, as estimated by our TI scheme and the
BAR method.  Figure~\ref{fig:contang_sw} shows that cosine of the
contact angle increases with the density of the structured wall,
with a sharp peak when the density of the wall is equal to the
average density of the bulk crystal. At this value of the wall
density, we found that the corresponding contact angle reaches
a minimum and is about $25^{\circ}$, which is in the partial
wetting regime. Contact angles are less than $90^{\circ}$ ($0<\cos
\theta_{\rm c}<1$) at all wall densities except at the smallest
$\rho_{\text{sw}}$ since there $\gamma_{\text{wc}}>\gamma_{\text{wl}}$
(see Fig.~\ref{fig:ifesw_vs_rhow}).

Figure~\ref{fig:contang_sw} also shows $\cos\theta_{\rm c}$ increasing
with increasing interaction strength between the the wall and the
LJ system.  This is because $\gamma_{\rm wl}$ and $\gamma_{\rm wc}$
gradually diverge from each other as $\epsilon_{\rm sw}$ increases,
as is clearly seen in Fig.~\ref{fig:ifesw_vs_ew}.  $0<\cos \theta_{\rm
c}<1$ at all values of $\epsilon_{\rm sw}$ indicating that the LJ crystal
partially wets the wall for such parameter values.

\section{Conclusion}
\label{sec_conc}
In this work, extensive MD simulations of a LJ system were performed
to compare different techniques for the calculation of wall-liquid and
wall-crystal interfacial free energies. We find that the equilibrium
thermodynamic integration (TI) method is in very good agreement with
non-equilibrium work techniques such as the Bennett acceptance ratio
(BAR) method. Within the restricted range of applicalbility, this also
holds for Gibbs-Cahn (GC) integration and the pressure anisotropy (PI)
method. This indicates the accuracy of the obtained estimates and also
validates the TI scheme, proposed in our earlier work.

While the GC method is computationally faster compared to the TI or
BAR methods, care has to be taken when the coexistence pressure and
temperature vary non-linearly along the melting curve since then more
number of points are needed to obtain reliable values for the interfacial
free energy. As a matter of fact, non-equilibrium work methods are not
more efficient compared to conventional TI. However, such methods are
helpful to check for any hysteresis in the TI scheme.

Using the obtained values of $\gamma_{\rm wl}$ and $\gamma_{\rm wc}$
and the values for $\gamma_{\rm cl}$ calculated previously using the
``cleaving potential method'' \cite{davidchack-laird03}, we applied
Young's equation to determine contact angles for flat and structured
walls.  In general, one observes smaller contact angles at higher
temperatures along the melting curve when the LJ system is in contact
with the flat wall. For the structured wall interfaces, it is found that
the density of the structured wall and the interaction strength of the
LJ particles with the wall significantly influence the contact angle
and the degree of wetting can be changed by tuning such parameters.
In particular, the contact angle exhibits a strongly non-monotonic
behavior with respect to the density of the structured wall, as it shows
a sharp minima when the density of the structured wall is equal to that
of the bulk crystal.

Both for the flat and the structured walls we have obtained the conditions
under which partial wetting is observed. Thus, the models proposed in
this work can now be used for the study of heterogeneous nucleation.

\begin{acknowledgments}
We acknowledge support by the Deutsche Forschungsgemeinschaft (DFG) 
under Grant Nos.~HO 2231/6-2 and HO 2231/6-3.
\end{acknowledgments}

\end{document}